# On the transfer of optical orbital angular momentum to matter

Kayn A. Forbes

*School of Chemistry, University of East Anglia, Norwich, Norfolk, NR4 7TJ, United Kingdom*

Email: k.forbes@uea.ac.uk

**Abstract**

The prevailing notion is that the orbital angular momentum (OAM) of an optical vortex can only be transferred to the internal degrees of freedom (i.e. electronic motion) of materials through electric quadrupole and higher-order multipole interactions. Here it is highlighted how this is an artefact of the paraxial approximation and that optical OAM can be transferred to electronic motion via dipole transitions under the correct experimental conditions where non-paraxial, longitudinal fields must be accounted for.

**Introduction**

Optical vortices are a class of structured laser beams which, amongst numerous interesting properties, possess an optical orbital angular momentum (OAM) [1,2]. In a paraxial description, the OAM of an optical vortex is $\ell\hbar$ per photon in the direction of propagation, where $\ell \in \mathbb{Z}$ is known as the topological charge [3]. The unique feature of an optical vortex is that it propagates with a helical-phase $\mathrm{e}^{i\ell\phi}$, and it is the gradient of this azimuthal phase which is responsible for the optical OAM. Optical vortices have been widely used in a plethora of applications, most commonly optical manipulation, information and communications, imaging, and quantum entanglement [4–6]. The vast majority of experimental studies have involved microparticles rather than nano-sized objects, however a flourishing recent application has been in atomic optics and molecular spectroscopies [7,8]. An important issue in both studies is how OAM of a vortex is transferred to the external (centre of mass motion) and internal (electronic motion) degrees of freedom of atoms, molecules, and materials in general.

Transfer of optical OAM to the centre of mass motion leads to light-induced mechanical effects, such as rotational motion of atoms about the beam axis [9]. The possibility of whether the optical OAM of an optical vortex can couple to the internal motion of matter, which rather than producing mechanical effects would yield internal electronic transitions, for example, has been a longstanding issue [10,11,7,8]. The first study concluded that only via electric quadrupole (or higher order multipole) interactions with the field could such a transfer occur [12,13]; OAM could only be transferred to the centre of mass during dipole excitations. This study looked specifically at paraxially propagating light, which in the terminology of Lax et al. [14] would be termed a zeroth-order, fully transverse field. Numerous further studies supported this initial conclusion [15–23], though some reported it was in fact possible to transfer optical OAM via dipole transitions with various mechanisms [24–26]. A key piece of work was the experimental proof from Schmiegelow et al. [27] that in a laser-cooled $^{40}$Ca$^+$ atom trapped using a Paul trap, the $4^2S_{1/2} \leftrightarrow 3^2D_{5/2}$ quadrupole transition rate is significantly modified when using an input Laguerre-Gaussian vortex beam compared to a non-OAM possessing Gaussian beam, indicating a transfer of OAM to the bound valence electron. Interestingly, it was soon realised that the experimental results presented in [27] can only be quantitatively understood if (non-paraxial) longitudinal fields were accounted for [28]. Of course, whilst the Schmiegelow et al. experiment verifies that OAM can be transferred to internal electronic motion in quadrupole transitions, it does not prove that it cannot be transferred in dipole interactions.

A particularly important aspect of OAM transfer to internal degrees of freedom is that the absorption and emission of photons can be sensitive to the *sign* (as well as magnitude) of optical OAM, through optical transition rates which exhibit a linear dependence on $\ell$. In other words, the rate of optical processes may be dependent on the handedness (or chirality) of the input optical vortex, analogous to the well-known differential couplings to circularly polarized light in optical activity [29,30] which stem from the spin angular momentum (SAM) of $\sigma\hbar$ per photon where $\sigma = \pm 1$ depending on the handedness of circular polarization.

In this letter the generic origin of how optical OAM can be transferred to the internal electronic motions of materials is highlighted: namely the transverse gradient of an optical vortex must be engaged in the light-matter coupling in order to exhibit effects dependent on the OAM. This fundamental result is of crucial importance for spectroscopic applications of twisted light, where the internal degrees of freedom of atoms and molecules play the significant role: as stated, centre of mass motions induced by the light correspond to mechanical effects. An application of this result shows how under specific circumstances the OAM of an optical vortex can in fact be transferred to matter via dipole transitions. Further, the difference between the two internal angular momenta transfers (SAM and OAM) are clarified with respective to the many recently observed chiroptical effects: SAM transfer probes the local helicity of circularly-polarized light, whereas $\ell$-dependent spectroscopic interactions are spatially-dependent phenomena.

**Interaction of an optical vortex with matter**

We use non-relativistic quantum electrodynamics (QED) as the theoretical framework in this analysis [31], specifically the Power-Zienau-Woolley (PZW) formulation [32,33]. The coupling between matter (a distinct unit of which, such as an atom or molecule, is denoted by $\xi$) and light in the PZW formulation is given by the interaction Hamiltonian in multipolar form:

$$H_{\text{int}}(\xi) = -\varepsilon_0^{-1}\mu_i(\xi)d_i^\perp(\mathbf{R}_\xi) - m_i(\xi)b_i(\mathbf{R}_\xi) - \varepsilon_0^{-1}Q_{ij}(\xi)\nabla_i d_j^\perp(\mathbf{R}_\xi) - ..., \tag{1}$$

where $\mu(\xi)$ is the electric dipole transition moment operator, with $m(\xi)$ and $Q_{ij}(\xi)$ the corresponding transition operators the magnetic dipole and electric quadrupole, respectively; $\mathbf{d}^\perp(\mathbf{R}_\xi)$ and $\mathbf{b}(\mathbf{R}_\xi)$ are the electric displacement field and magnetic field operators, respectively (both transverse to the Poynting vector and not neccesarily the direction of beam propagation); we use standard suffix notation for tensor quantities and imply the Einstein summation convention for repeated indices throughout (i.e. $a_i b_i = \mathbf{a} \cdot \mathbf{b}$). In (1) the centre of mass motion is neglected, as is standard, because for the majority of situations the Born-Oppenheimer approximation is valid, and thus the dynamics associated with (1) are with respect to the internal electronic coordinates.

The electromagnetic field operators for the Laguerre-Gaussian optical vortex modes were previously derived in ref [34], and, the terminogloy of Lax *et al.*, only included the zeroth-order and fully transverse [14,35] (with respect to the propagation direction) component of the field – we now denote these fields $T_0$. However, although longitudinal components to the field are well known to be important for non-paraxial and strongly-focussed light [36], it has recently been highlighted how the first-order longitudinal components (denoted $L_1$) of such fields are in fact requisite even for beams that would be considered propagating within the paraxial regime [37]. The mode expansion for circularly-polarized LG beams that include both the $T_0$ and $L_1$ components are given as

$$d_i^\perp(\mathbf{r}) = \sum_{k,\sigma,\ell,p} \Omega \left[ \left\{ (\hat{x} + i\sigma\hat{y})_i + \frac{i}{k}\left(\frac{\partial}{\partial r} - \frac{\ell\sigma}{r}\right) e^{i\sigma\phi} \hat{z}_i \right\} f_{|\ell|,p}(r) a_{|\ell|,p}^{(\sigma)}(k\hat{z}) e^{i(kz+\ell\phi)} - H.c \right], \qquad (2)$$

and

$$b_i(\mathbf{r}) = \sum_{k,\sigma,\ell,p} \frac{\Omega}{c} \left[ \left\{ (\hat{y} - i\sigma\hat{x})_i + \frac{1}{k}\left(\sigma\frac{\partial}{\partial r} - \frac{\ell}{r}\right) e^{i\sigma\phi} \hat{z}_i \right\} f_{|\ell|,p}(r) a_{|\ell|,p}^{(\sigma)}(k\hat{z}) e^{i(kz+\ell\phi)} - H.c \right], \qquad (3)$$

where $\Omega = i\left(\hbar c k \varepsilon_0 / 4 A_{\ell,p}^2 V\right)^{1/2}$ is the normalization constant for LG modes, $\sqrt{2}^{-1}(\hat{x}(\hat{y}) \pm i\sigma\hat{y}(\hat{x}))_i$ represents the electric (magnetic) polarization vector for circularly-polarized light, with $V$ the quantization volume; $f_{|\ell|,p}(r)$ is a radial distribution function; $a_{|\ell|,p}^{(\sigma)}(k\hat{z})$ is the annihilation operator; and *H.c.* stands for Hermitian conjugate. The terms that depend on $\hat{x}$ and $\hat{y}$ are the $T_0$ parts and those that depend on $\hat{z}$ are the $L_1$.

**$T_0$ interactions**

Looking at the interaction Hamiltonian (1), the first term is the electric-dipole coupling term (E1), the second term is the magnetic-dipole (M1) interaction and the third term is the electric-quadrupole term (E2). The most clear and obvious difference between the three is that the E2 term depends on the gradient of the electric field through the $\nabla d^\perp(\mathbf{r})$ term, whilst the other terms represent direct coupling between the transition moment operators and the respective electric and magnetic field operator.

An important and subtle point that must be expanded upon for reference later is that it can be said that the magnetic field comes from the gradient of the electric field, as most clearly seen in Faraday's law $\nabla \times e^\perp(\mathbf{r}) = -\partial b(\mathbf{r})/\partial t$. However, for a $T_0$ electric field and magnetic field, the ensuing cross product gives terms for a magnetic field that stem from $\partial/\partial z$, i.e. *longitudinal phase* gradients.

It is clear to see then that for a $T_0$ description of an LG mode, neither E1 or M1 couplings are linearly dependent on $\ell$ - i.e. the OAM of the light. However, the electric quadrupole dependence on the gradient of the field is given by:

$$Q_{ij}\nabla_j d_{i(T_0)}^{\perp} \propto Q_{ij}\nabla_j \hat{e}_i f_{\ell,p}(r)e^{(ikz+i\ell\phi)} = Q_{ij}\hat{e}_i\left[\hat{r}_j\partial_r + i\ell r^{-1}\hat{\phi}_j + ik\hat{z}_j\right]f_{\ell,p}(r)e^{(ikz+i\ell\phi)}. \quad (4)$$

And thus we see in the middle term dependent on $\ell$ that E2 interactions exhibit the ability to depend on the OAM of the input $T_0$ LG beam [17]. Of course, higher-order multipole interactions will also exhibit this effect, but E2 is the lowest order and thus most experimentally viable in general. The linear dependence on $\ell$ which stems from the E2 coupling being proportional to the azimuthal gradient of $e^{i\ell\phi}$ (it is crucial to distinguish $\ell$ from $|\ell|$) in the matrix element $M_{fi}=\langle f|Q_{ij}\nabla_j d_i^{\perp}|i\rangle$ is what indicates a transfer of OAM to the internal electronic coordinates.

**L$_1$ Interactions**

Certain $\hat{z}$-dependent L$_1$ terms in both (2) and (3) much more clearly exhibit their potential to engage the OAM through $\ell$ in interactions with matter. Importantly, it can be seen that even in dipole coupling this is the case: both $M_{fi}=\langle f|\boldsymbol{\mu}\cdot\boldsymbol{d}^{\perp}|i\rangle$ and $M_{fi}=\langle f|\boldsymbol{m}\cdot\boldsymbol{b}|i\rangle$ can exhibit terms linearly proportional to $\ell$.

The origins of the L$_1$ terms in the electric and magnetic fields are of utmost importance in understanding why this is the case: they can be determined from Gauss's law $\nabla\cdot\boldsymbol{d}=0$ and $\nabla\cdot\boldsymbol{b}=0$, respectively. In general, therefore:

$$\int \nabla^{\perp}\cdot\boldsymbol{d}_{x,y}^{\perp}(\boldsymbol{r})dz \propto \frac{i}{k}\left(\frac{\partial}{\partial x}+i\sigma\frac{\partial}{\partial y}\right)\boldsymbol{d}_{x,y}^{\perp}(\boldsymbol{r}) = d_z^{\perp}(\boldsymbol{r});$$

$$\int \nabla^{\perp}\cdot\boldsymbol{b}_{x,y}^{\perp}(\boldsymbol{r})dz \propto \frac{i}{k}\left(\frac{\partial}{\partial y}-i\sigma\frac{\partial}{\partial x}\right)\boldsymbol{b}_{x,y}(\boldsymbol{r}) = b_z(\boldsymbol{r}). \quad (5)$$

Clearly then both the L$_1$ components $d_z^{\perp}(r)$ and $b_z(r)$ stem from the *transverse* gradients of the $T_0$ fields, and the well-known coordinate transformations show that $\partial/\partial x$ and $\partial/\partial y$ are proportional to $\partial/\partial \phi$. It is now appropriate to remember the point made about the $T_0$ magnetic field and the fact it comes from the *longitudinal* gradient $\partial/\partial z$ of $T_0$ electric field.

So although distinct, and important under different experimental situations, the $\ell$-dependent interactions stemming from $T_0$ and $L_1$ parts of the LG modes are in fact related by the fact they both stem from the *transverse phase gradient* of the field. For LG modes (or any optical vortex mode) this involves a helical-phase gradient, which is in fact the defining feature of optical vortex modes as it is what gives them their OAM [3]. Although this fundamental result has not been previously pointed out, it should not come as a surprise that in order to engage and transfer the OAM of optical vortices in light-matter interactions and electronic motion, respectively, there must be a coupling to the helical-phase gradient. We have shown it may occur via $T_0$ fields with electric quadrupole or higher multipole interactions, or via $L_1$ fields in the dipole approximation. The importance of either route is dictated by experimental conditions such as the nature and orientation of the material itself as well as the optical parameters of the input mode (e.g. the polarization and topological charge) and the size of the paraxial parameter $kw_0$, where $k = 2\pi/\lambda$ is the wave number and $w_0$ the beam waist. For example, in the experimental study by Giammanco *et al.* [38] the electric dipole transitions they looked at showed no dependence on the input OAM, which is to be expected if looking at $T_0$ fields alone, but their value of $kw_0 \approx 20000$ means the longitudinal $L_1$ fields which are weighted by the factor $(kw_0)^{-1}$ are insignificant in their experiments.

**Analysis and Discussion**

There are some intricate points that must now be extracted from our results so far for the $L_1$ fields. For $L_1$ E1 matrix elements, if $\sigma = 0$ there is no potential for the transfer of OAM via an electric-dipole interaction as any terms dependent on $\ell$ vanish in (2). For cases where the light is circularly-polarized and $\sigma = \pm 1$ it becomes important whether the OAM and SAM are parallel or anti-parallel, that is $\text{sgn}\,\ell = \text{sgn}\,\sigma$ and $\text{sgn}\,\ell = -\text{sgn}\,\sigma$, respectively. In the parallel case the total angular momentum of the photon is clearly $J_{z(\text{photon})} = (|\ell|+1)\hbar$ whereas in the latter case it is $J_{z(\text{photon})} = (|\ell|-1)\hbar$. Calculating the $L_1$ contribution to the matrix element for the important case of $p = 0$ for the indicative process of single photon absorption yields:

$$M_{fi}^{L_1 E_1} \propto \frac{i}{k}\left(\frac{\partial}{\partial r} - \ell\sigma\frac{1}{r}\right)e^{i(\ell+\sigma)\phi} f_{|\ell|,0}(r)\hat{z}\cdot\boldsymbol{\mu}^{\alpha 0} \propto \frac{i}{kr}\left(|\ell| - \ell\sigma\right)e^{i(\ell+\sigma)\phi} f_{|\ell|,0}(r)\hat{z}\cdot\boldsymbol{\mu}^{\alpha 0}, \qquad (6)$$

thus the $|\ell|$ term always cancels out the $\ell\sigma$ term when $\text{sgn}\,\ell = \text{sgn}\,\sigma$, and so any $\ell$-dependent coupling (OAM transfer) is always zero in the parallel case. It is non-zero in general for the anti-parallel case. However we now come to an important point with regards to the separation of orbital and spin angular momentum of photons. When dealing with $T_0$ fields the separation of SAM and OAM of light is well-defined: each photon may possess discrete values of $\sigma\hbar$ of SAM and $\ell\hbar$ of OAM in this paraxial regime along the direction of propagation [39]. For linearly polarized light $\sigma = 0$ and the paraxial result of $\ell\hbar$ OAM is in fact exact and carries over to non-paraxial situations. In either case, the total angular momentum $J_{z(\text{photon})}$ per photon is consistently $(\ell+\sigma)\hbar$ and previous theory and experiments working *within the paraxial regime* have been able to categorically state that OAM transfer to electrons does not occur through dipole interactions, but must engage E2 or higher couplings [7] (as also pointed out in here in the *$T_0$ interactions* section). This is only possible because any discussion of angular momentum conservation laws must require the angular momenta concerned to be well-defined themselves [40] – for $T_0$ fields both the SAM and OAM of photons are well-defined in discrete values of $\hbar$; for total fields that include $L_1$ possessing SAM they are not.

Although it has been shown that L$_1$ fields are exhibited in light that can be considered to be propagating paraxially [37], it is clear that the SAM and OAM are mixing through a spin-orbit interaction of light (SOI) in the L$_1$ E1 case, most usually encountered in non-paraxial beams [36], where the partitioning of SAM and OAM into $\sigma\hbar$ and $\ell\hbar$ has long been known to be problematic [39,41,42]. Put more succinctly, inclusion of non-paraxial L$_1$ fields means that the simple separation of angular momentum into $\sigma$-dependent spin and $\ell$-dependent orbital components in units of $\hbar$ is not possible. So although it is incorrect to state that discrete $\ell\hbar$ values of OAM can be transferred in electric dipole transitions, we have highlighted here that a degree of OAM can be transferred in electric dipole transitions if the input beam has $\sigma \neq 0$. The expectation value of this OAM can be determined by known equations, e.g. those found in [43,44].

The scenario is different for L$_1$ M1 interactions, where the same situation discussed in the previous paragraph arises if $\sigma = \pm 1$, but crucially if the vortex is linearly-polarized and possesses no SAM ($\sigma = 0$), there still exists a non-zero matrix element proportional to $\ell$ which thus indicates the transfer of discrete $\ell\hbar$ optical OAM. The matrix element for single-photon absorption through a magnetic dipole transition with the L$_1$ component of the magnetic field is given by

$$M_{fi}^{L_1 M_1} \propto \frac{1}{k}\left(\sigma \frac{\partial}{\partial r} - \frac{\ell}{r}\right) e^{i(\ell+\sigma)\phi} f_{|\ell|,0}(r) \hat{z} \cdot \boldsymbol{m}^{\alpha 0} \propto \frac{1}{kr}\left(\sigma|\ell| - \ell\right) e^{i(\ell+\sigma)\phi} f_{|\ell|,0}(r) \hat{z} \cdot \boldsymbol{m}^{\alpha 0}. \qquad (7)$$

As mentioned, even though the orbital and spin angular momentum are not well-defined for non-paraxial fields in general, in the case of when $\sigma = 0$, the $\ell\hbar$ of OAM per photon is correct in either paraxial or non-paraxial optics. As such, we can state categorically that $\ell\hbar$ optical OAM can be transferred in magnetic dipole transitions.

Interestingly, even though the vast majority of OAM-dependent spectroscopic effects have been due to the electric polarization (electric multipole moments e.g. electric dipole and electric quadrupole) of the material coupling to the optical vortex, some previous studies have stated and observed such effects stemming from magnetic dipole interactions [45–48]. Given the results above it is of no surprise that in all of those works the key factor is the non-paraxality of the fields used, and thus the importance of significant L$_1$ contributions to the total field. Finally, the magnitudes of matrix elements involving structured laser light (including those discussed in this section) are highly-dependent on where the absorbing particle is within the transverse beam profile. This position-dependence has been well studied e.g. [49–52], and alongside well-known selection rules governs the magnitude of any optical coupling.

*Optical chirality and optical helicity*

One of the key consequences (or applications) of the ability to transfer optical OAM to electronic motion has been the ability to engage the chirality of optical vortices in light-matter interactions. That is, in certain cases of where OAM is transferred, the resultant optical transitions rates (which are proportional to the modulus square of the matrix element) for absorption or scattering processes are then linearly-dependent on $\ell$, i.e. the handedness of the input optical vortex.

The well-defined quantities known as optical chirality and optical helicity play an important role in chiral light-matter interactions and optical activity [53,54]. The most well-known exhibition is from circularly-polarized light interacting differentially with chiral matter: simply put, spectroscopic optical processes, such as absorption and scattering, are dependent on the value of $\sigma$, itself determined by whether the input beam is right CPL or left CPL [29]. Naturally one can pose the question of whether optical interactions (specifically non-mechanical) can be dependent on the sign of $\ell$, i.e. whether the optical vortex twists to the right or to the left. This field of chirality and the optical OAM of light has recently been surveyed in reference [8]. Going back to the well-defined quantities of optical chirality and optical helicity, it is important to realise that these specifically relate to the local property of polarization (and helicity (SAM)); optical OAM is a spatial property of light, and so there is no way to universally quantify or measure absolutely any chirality that is associated with it. This is no different to trying to measure the spatial (geometrical) chirality of a molecule – there is no quantum operator for spatial chirality, and so it is not absolutely measureable and thus not quantifiable. For example, a recent experimental study has most clearly highlighted this idea by using macroscopic chiral particles which exhibit a peak chiral interaction with an optical vortex when $\ell \sim \pm 32$, either side of this maximum the interaction decreases (i.e. the beam's 'degree of chirality' constantly changes depending on the system it is interacting with) [55]. Therefore a clear distinction is made with respect to chiroptical effects using circular-polarization or optical vortices – the former probes the local helicity of light and the position of a molecule is irrelevant, the latter relies on probing the spatial chirality of the light. Unlike effects dependent solely on the chirality associated with polarization, any chiroptical or optical activity that depends on $\ell$ (including $\ell\sigma$) is an acutely spatially-dependent phenomenon.

**Conclusion**

Stating that the OAM of light may only be transferred to internal degrees of freedom via quadrupole transitions is an artefact that originates from the paraxial approximation. Beyond the paraxial approximation, the inclusion of longitudinal $L_1$ fields provides a means for the OAM of an optical vortex to be transferred to an electron via dipole transitions. Specifically, whilst SOI and the non-discrete nature of SAM and OAM for non-paraxial fields means that it cannot be stated that discrete units of $\ell\hbar$ of OAM is transferred in electric dipole transitions, a degree of OAM can be transferred. An analogous result applies to magnetic dipole interactions when the light possesses SAM, in contrast to electric dipole transitions however, OAM may be transferred as $\ell\hbar$ to matter via magnetic dipole transitions in cases where $\sigma = 0$. The relevance of either method of OAM transfer is dictated by the material interacting with the optical vortex, alongside the optical angular momentum quantities of the mode and degree of focussing. We have clarified the unified origin of both distinct phenomena being in the dependence on the transverse phase gradient (i.e. the azimuthal or helical phase gradient) of the optical vortex, a seemingly obvious but hitherto unrealized fact. The ability for OAM to influence optical phenomena in the dominant dipole regime rather than requiring small quadrupole interactions significantly opens up the usefulness and applicability of twisted light in non-mechanical light-matter interactions.

Furthermore, the transfer and engagement of optical OAM in matrix elements allows for optical transition rates to be chirally sensitive to the vortex handedness through a linear dependence on $\ell$. Any such effect is acutely spatially-dependent, unlike standard optical activity stemming from SAM and $\sigma$ which probe the local helicity of light. Being a spatial effect means that these phenomena cannot be specifically quantified, as they relate to geometrically chirality of which no quantum operator exists.

**Acknowledgements**

KAF is grateful to David L. Andrews for helpful discussions and to the Leverhulme Trust for funding him through a Leverhulme Trust Early Career Fellowship (Grant Number ECF-2019-398).

**References**

1. D. L. Andrews, ed., *Structured Light and Its Applications: An Introduction to Phase-Structured Beams and Nanoscale Optical Forces* (Academic press, 2011).
2. D. L. Andrews and M. Babiker, eds., *The Angular Momentum of Light* (Cambridge University Press, 2012).
3. L. Allen, M. W. Beijersbergen, R. J. C. Spreeuw, and J. P. Woerdman, "Orbital angular momentum of light and the transformation of Laguerre-Gaussian laser modes," Phys. Rev. A **45**(11), 8185–8189 (1992).
4. A. M. Yao and M. J. Padgett, "Orbital angular momentum: Origins, behavior and applications," Adv. Opt. Photonics **3**(2), 161–204 (2011).
5. M. J. Padgett, "Orbital angular momentum 25 years on [Invited]," Opt. Express **25**(10), 11265–11274 (2017).
6. S. M. Barnett, M. Babiker, and M. J. Padgett, eds., *Optical Orbital Angular Momentum* (The Royal Society, 2017).
7. M. Babiker, D. L. Andrews, and V. E. Lembessis, "Atoms in complex twisted light," J. Opt. **21**(1), 013001 (2018).
8. K. A. Forbes and D. L. Andrews, "Orbital angular momentum of twisted light: chirality and optical activity," J. Phys. Photonics (2021).
9. M. F. Andersen, C. Ryu, P. Cladé, V. Natarajan, A. Vaziri, K. Helmerson, and W. D. Phillips, "Quantized rotation of atoms from photons with orbital angular momentum," Phys. Rev. Lett. **97**(17), 170406 (2006).
10. S. J. Van Enk, "Selection rules and centre-of-mass motion of ultracold atoms," Quantum Opt. J. Eur. Opt. Soc. Part B **6**(5), 445 (1994).
11. S. Franke-Arnold, "Optical angular momentum and atoms," Phil Trans R Soc A **375**(2087), 20150435 (2017).
12. M. Babiker, C. R. Bennett, D. L. Andrews, and L. D. Romero, "Orbital angular momentum exchange in the interaction of twisted light with molecules," Phys. Rev. Lett. **89**(14), 143601 (2002).
13. S. M. Lloyd, M. Babiker, and J. Yuan, "Interaction of electron vortices and optical vortices with matter and processes of orbital angular momentum exchange," Phys. Rev. A **86**(2), 023816 (2012).
14. M. Lax, W. H. Louisell, and W. B. McKnight, "From Maxwell to paraxial wave optics," Phys. Rev. A **11**(4), 1365 (1975).
15. V. E. Lembessis and M. Babiker, "Enhanced quadrupole effects for atoms in optical vortices," Phys. Rev. Lett. **110**(8), 083002 (2013).
16. A. Afanasev, C. E. Carlson, and A. Mukherjee, "High-multipole excitations of hydrogen-like atoms by twisted photons near a phase singularity," J. Opt. **18**(7), 074013 (2016).
17. K. A. Forbes and D. L. Andrews, "Optical orbital angular momentum: twisted light and chirality," Opt. Lett. **43**(3), 435–438 (2018).
18. T. Arikawa, T. Hiraoka, S. Morimoto, F. Blanchard, S. Tani, T. Tanaka, K. Sakai, H. Kitajima, K. Sasaki, and K. Tanaka, "Transfer of orbital angular momentum of light to plasmonic excitations in metamaterials," Sci. Adv. **6**(24), eaay1977 (2020).

19. W. Brullot, M. K. Vanbel, T. Swusten, and T. Verbiest, "Resolving enantiomers using the optical angular momentum of twisted light," Sci. Adv. **2**(3), e1501349 (2016).
20. M. van Veenendaal and I. McNulty, "Prediction of strong dichroism induced by x rays carrying orbital momentum," Phys. Rev. Lett. **98**(15), 157401 (2007).
21. R. Mathevet, B. V. de Lesegno, L. Pruvost, and G. L. Rikken, "Negative experimental evidence for magneto-orbital dichroism," Opt. Express **21**(4), 3941–3945 (2013).
22. A. A. Peshkov, D. Seipt, A. Surzhykov, and S. Fritzsche, "Photoexcitation of atoms by Laguerre-Gaussian beams," Phys. Rev. A **96**(2), 023407 (2017).
23. S. Bougouffa and M. Babiker, "Quadrupole absorption rate and orbital angular momentum transfer for atoms in optical vortices," Phys. Rev. A **102**(6), 063706 (2020).
24. A. Alexandrescu, D. Cojoc, and E. Di Fabrizio, "Mechanism of angular momentum exchange between molecules and Laguerre-Gaussian beams," Phys. Rev. Lett. **96**(24), 243001 (2006).
25. P. K. Mondal, B. Deb, and S. Majumder, "Angular momentum transfer in interaction of Laguerre-Gaussian beams with atoms and molecules," Phys. Rev. A **89**(6), 063418 (2014).
26. A. Bhowmik, P. K. Mondal, S. Majumder, and B. Deb, "Interaction of atom with nonparaxial Laguerre-Gaussian beam: Forming superposition of vortex states in Bose-Einstein condensates," Phys. Rev. A **93**(6), 063852 (2016).
27. C. T. Schmiegelow, J. Schulz, H. Kaufmann, T. Ruster, U. G. Poschinger, and F. Schmidt-Kaler, "Transfer of optical orbital angular momentum to a bound electron," Nat. Commun. **7**, 12998 (2016).
28. G. F. Quinteiro, F. Schmidt-Kaler, and C. T. Schmiegelow, "Twisted-light–ion interaction: the role of longitudinal fields," Phys. Rev. Lett. **119**(25), 253203 (2017).
29. L. D. Barron, *Molecular Light Scattering and Optical Activity* (Cambridge University Press, 2009).
30. M. Krupová, J. Kessler, and P. Bour, "Recent Trends in Chiroptical Spectroscopy: Theory and Applications of Vibrational Circular Dichroism and Raman Optical Activity," ChemPlusChem (2020).
31. D. P. Craig and T. Thirunamachandran, *Molecular Quantum Electrodynamics: An Introduction to Radiation-Molecule Interactions* (Courier Corporation, 1998).
32. D. L. Andrews, G. A. Jones, A. Salam, and R. G. Woolley, "Perspective: Quantum Hamiltonians for optical interactions," J. Chem. Phys. **148**(4), 040901 (2018).
33. D. L. Andrews, D. S. Bradshaw, K. A. Forbes, and A. Salam, "Quantum electrodynamics in modern optics and photonics: tutorial," JOSA B **37**(4), 1153–1172 (2020).
34. L. D. Romero, D. L. Andrews, and M. Babiker, "A quantum electrodynamics framework for the nonlinear optics of twisted beams," J. Opt. B Quantum Semiclassical Opt. **4**(2), S66–S72 (2002).
35. L. W. Davis, "Theory of electromagnetic beams," Phys. Rev. A **19**(3), 1177 (1979).
36. K. Y. Bliokh, F. J. Rodríguez-Fortuño, F. Nori, and A. V. Zayats, "Spin–orbit interactions of light," Nat. Photonics **9**(12), 796–808 (2015).
37. K. A. Forbes, D. Green, and G. A. Jones, "Relevance of Longitudinal Fields of Paraxial Optical Vortices," ArXiv201209555 Phys. (2020).
38. F. Giammanco, A. Perona, P. Marsili, F. Conti, F. Fidecaro, S. Gozzini, and A. Lucchesini, "Influence of the photon orbital angular momentum on electric dipole transitions: negative experimental evidence," Opt. Lett. **42**(2), 219–222 (2017).
39. S. M. Barnett and L. Allen, "Orbital angular momentum and nonparaxial light beams," Opt. Commun. **110**(5–6), 670–678 (1994).
40. S. J. Van Enk and G. Nienhuis, "Commutation rules and eigenvalues of spin and orbital angular momentum of radiation fields," J. Mod. Opt. **41**(5), 963–977 (1994).
41. S. M. Barnett, "Optical angular-momentum flux," J. Opt. B Quantum Semiclassical Opt. **4**(2), S7 (2001).
42. S. J. Van Enk and G. Nienhuis, "Spin and orbital angular momentum of photons," EPL Europhys. Lett. **25**(7), 497 (1994).

43. K. Y. Bliokh, J. Dressel, and F. Nori, "Conservation of the spin and orbital angular momenta in electromagnetism," New J. Phys. **16**(9), 093037 (2014).
44. K. Y. Bliokh, E. A. Ostrovskaya, M. A. Alonso, O. G. Rodríguez-Herrera, D. Lara, and C. Dainty, "Spin-to-orbital angular momentum conversion in focusing, scattering, and imaging systems," Opt. Express **19**(27), 26132–26149 (2011).
45. T. Wu, R. Wang, and X. Zhang, "Plasmon-induced strong interaction between chiral molecules and orbital angular momentum of light," Sci. Rep. **5**, 18003 (2015).
46. Y. Guo, G. Zhu, W. Bian, B. Dong, and Y. Fang, "Orbital angular momentum dichroism caused by the interaction of electric and magnetic dipole moments and the geometrical asymmetry of chiral metal nanoparticles," Phys. Rev. A **102**(3), 033525 (2020).
47. P. Woźniak, I. D. Leon, K. Höflich, G. Leuchs, and P. Banzer, "Interaction of light carrying orbital angular momentum with a chiral dipolar scatterer," Optica **6**(8), 961–965 (2019).
48. C. Rosales-Guzmán, K. Volke-Sepulveda, and J. P. Torres, "Light with enhanced optical chirality," Opt. Lett. **37**(17), 3486–3488 (2012).
49. A. Afanasev, C. E. Carlson, and A. Mukherjee, "Off-axis excitation of hydrogenlike atoms by twisted photons," Phys. Rev. A **88**(3), 033841 (2013).
50. A. Afanasev, C. E. Carlson, C. T. Schmiegelow, J. Schulz, F. Schmidt-Kaler, and M. Solyanik, "Experimental verification of position-dependent angular-momentum selection rules for absorption of twisted light by a bound electron," New J. Phys. **20**(2), 023032 (2018).
51. Y. Duan, R. A. Müller, and A. Surzhykov, "Selection rules for atomic excitation by twisted light," J. Phys. B At. Mol. Opt. Phys. **52**(18), 184002 (2019).
52. V. V. Klimov, D. Bloch, M. Ducloy, and J. R. Leite, "Mapping of focused Laguerre-Gauss beams: The interplay between spin and orbital angular momentum and its dependence on detector characteristics," Phys. Rev. A **85**(5), 053834 (2012).
53. M. M. Coles and D. L. Andrews, "Chirality and angular momentum in optical radiation," Phys. Rev. A **85**(6), 063810 (2012).
54. F. Crimin, N. Mackinnon, J. B. Götte, and S. M. Barnett, "Optical helicity and chirality: conservation and sources," Appl. Sci. **9**(5), 828 (2019).
55. J. Ni, S. Liu, D. Wu, Z. Lao, Z. Wang, K. Huang, S. Ji, J. Li, Z. Huang, Q. Xiong, Y. Hu, J. Chu, and C.-W. Qiu, "Gigantic vortical differential scattering as a monochromatic probe for multiscale chiral structures," Proc. Natl. Acad. Sci. **118**(2), e2020055118 (2021).